\documentclass[review]{elsarticle}

\usepackage{lineno,hyperref,amsmath, amssymb}

\biboptions{numbers,sort&compress}

\modulolinenumbers[5]

\journal{Journal of \LaTeX\ Templates}

\begin{document}

\begin{frontmatter}

\title{Squeezed Coherent States of Motion for Ions Confined in Quadrupole and Octupole Ion Traps}

\author{Bogdan M. Mihalcea}
\address{Natl. Inst. for Laser, Plasma and Radiation Physics (INFLPR), \\ Atomi\c stilor Str. Nr. 409, 077125 M\u agurele, Romania}
\ead{bogdan.mihalcea@inflpr.ro}




\begin{abstract}
Quasiclassical dynamics of trapped ions is characterized by applying the time dependent variational principle (TDVP) on coherent state orbits, in case of quadrupole and octupole combined (Paul and Penning) and radiofrequency (RF) traps. A dequantization algorithm is proposed, by which the classical Hamilton (energy) function associated to the system results as the expectation value of the quantum Hamiltonian on squeezed coherent states. We develop such method and particularize the quantum Hamiltonian for a combined and for a RF trap, with axial symmetry and a RF anharmonic electric potential. We also build the classical Hamiltonian functions for the particular traps we considered, and find the classical equations of motion.   
\end{abstract}


\end{frontmatter}


\section{Introduction}\label{intro}

Recent advances in quantum optics \cite{Orszag2016} enable trapping of single particles or atoms \cite{Blaum2008, Haroche2013}, while progress in quantum engineering techniques allows preparing these particles in well-defined quantum states \cite{Leibf2003, Wine2013, Knoop2014}. Quantum engineering using ion traps offers the possibility to prepare stable quantum states by precise control of the interactions between a quantum system (trapped ions) and the environment \cite{Sinclair2011, Blatt2012, Knoop2015}, while investigation of nonclassical states of spin systems coupled to a harmonic oscillator offers the possibility to investigate fundamental quantum phenomena, such as the mechanisms responsible for decoherence and the quantum-classical transition \cite{Sawyer2012, Lo2015}. 

The quantum time-dependent harmonic oscillator has been intensively used to describe the dynamics of many physical systems \cite{Combe86, Combe87b, Combe88, Davydov2011}. Quantum dynamics of harmonic oscillators is obtained by the use of the so-called {\em Peremolov's generalized coherent states} \cite{Perel86, Combe92} of the Lie algebra associated to the SU(1,1) group \cite{Combe2012}. Quantum  dynamics in a 3D RF ion trap characterized by a uniform magnetic field and a time-dependent quadrupole electric potential, can be reduced to the solution of the time-dependent quantum oscillator equations, obtained by separating the axial and radial motion from the Schr\"odinger equation \cite{Gheor92}. In addition, the coherent state approach leads to quantum solutions that are explicitly constructed as functions of the classical trajectories on the phase space \cite{Combe92}. The properties of a RF (Paul)-trap with a super-imposed magnetic field (combined trap) are presented in \cite{Li1992}, and it was shown that the regions of stability are significantly larger than those for a Paul trap for both positive and negative charged ions. Ion dynamics in a radiofrequency (RF) octupole trap was described in \cite{Walz1993, Mihalcea1999}, demonstrating confinement of ions in a RF anharmonic electric potential and characterizing the stability of this nonlinear parametric oscillator. Collective dynamics for systems of ions confined in quadrupole 3D traps with cylindrical symmetry is characterized in \cite{Gheor2000}. Ion dynamics in a linear combined trap has been investigated both theoretically and experimentally in \cite{Nakamura2001}, and it was shown that the presence of a homogeneous magnetic field superimposed over the applied DC and RF electric fields, leads to a set of coupled Mathieu equations. Coherent states for a set of quadratic Hamiltonians in the trap regime are constructed in Ref. \cite{Astorga2010}, and then particularized to the asymmetric Penning trap. A method of finding a set of generators that form a closed Lie algebra, which creates a framework to characterize a general quantum Hamiltonian is presented in \cite{Ibarra2015}. Thus, the Lie algebra can be extended to study the Hamiltonian of a bi-dimensional charged particle in time-dependent electromagnetic fields, by exploiting the similarities between the terms of these two Hamiltonians. 

This paper characterizes the evolution of squeezed coherent states of motion for ions confined in quadrupole and octupole combined and RF traps, using the coherent state formalism \cite{Klaud85, Perel86, Gheor92} developed in \cite{Major2005, Combe2012}, and the time dependent variational principle (TDVP) \cite{Krame81}. We study the bosonic realization of the Lie algebra for the SU(1,1) group, and (generalized) coherent states in the Fock space for a trapped ion \cite{Nieto2000}. The paper is organized as follows: In Section \ref{dynsym} we propose a dequantization algorithm \cite{Abrikosov2005} that enables explicit calculus of the quantum Hamilton function associated to an anharmonic oscillator (ion) confined in a combined or RF trap, which describes an algebraic model when the anharmonic part is a polynomial. Such model is linear for 3D quadrupole ion traps that exhibit axial symmetry. Section \ref{energy} presents the solutions of the Schr{\"o}dinger equation and the quasienergy spectrum for the model we suggest. The method suggested in \cite{Gheor92, Mihalcea2011} is developed and particularized in Section \ref{quasi} for both combined (Paul and Penning) and RF traps, considering a RF anharmonic electric potential. We build the classical Hamilton functions for such particular traps and find the classical Hamilton equations of motion for the anharmonic combined trap, which represents another original result. We suggest that in the pseudopotential approximation case (ideal RF trap), the points of minimum of the classical Hamiltonian describe equilibrium configurations for trapped ions, of interest for implementing quantum logic. The results are discussed in Section \ref{Concl}, as the model is straightforward to extend for 2D ion traps.

\section{Quantum dynamics of ions in axially symmetric quadrupole and octupole ion traps}
\paragraph{Algebraic models for axially symmetric nonlinear quadrupole and octupole traps}
\label{dynsym}

The dynamical group $Sp\left(2, {\mathbb R}\right)_a\otimes Sp\left(2, {\mathbb R}\right)_r\otimes SO\left(2\right)$ associated to the Hamilton function which describes the dynamics of an ion of mass $M$ and electric charge $Q$, confined within a 3D RF ion trap that exhibits both cylindrical and reflection symmetry, is the direct product between the axial and radial symplectic groups, and the rotations group $SO\left(2\right)$ generated by the axial angular moment operator $L_z$ \cite{Major2005}. The Lie algebra basis of $Sp\left(2, {\mathbb R}\right)_j , \;j = a, r$, is spanned by the generators $K_{0j}, K_{1j}$ and $K_{2j}$. We introduce the infinitesimal generators of the axial symplectic group $Sp\left(2, {\mathbb R}\right)_a$ defined as \cite{Major2005, Gheor92}
\begin{equation}\label{quadyn35}
K_{0,1a} = \frac{M\omega_a}{4\hbar}\left[z^2 \pm \frac{p_z^2}{M^2\omega_a^2}\right] \ , \;
K_{2a} = \frac {i}{4\hbar} \left[2z\frac \partial{\partial z} + 1\right] \ ,
\end{equation}
where $\omega_a/2 \pi$ is the trap axial frequency. The commuting relations are
\begin{equation}\label{quadyn36}
\left[K_{0a}, K_{1a}\right] = i K_{2a} \ , \;
\left[K_{2a}, K_{0a}\right] = iK_{1a} \ , \;  \left[K_{1a}, K_{2a}\right] = iK_{0a} \ .
\end{equation}
Using eqs. (\ref{quadyn35}) we infer 
\begin{equation}\label{quadyn37}
\left(K_{0a} + K_{1a}\right) \left(K_{0a} - K_{1a}\right) = iK_{2a} + \frac 1{4\hbar^2}z^2 p_z^2  \ .
\end{equation}
The Casimir operator that determines the Bargmann indices $k$ is \cite{Major2005, Perel86, Gheor92}
\begin{equation}
\label{quadyn41}C_{2a} = K_{0a}^2 - K_{1a}^2 - K_{2a}^2 = - \frac 3{16}I = k\left(k-1\right) I \ .
\end{equation}
where $I$ is the unit operator. The Bargmann indices that characterize axial motion are $k_{a+} = \frac 14\ ,\;k_{a-} = \frac 34$. We use cylindrical coordinates $\rho$ and $\theta$, with radial coordinates $x = \rho \cos \theta$ and $y = \rho \sin \theta$. The radial symplectic group generators for fixed angular moment $l$ are \cite{Major2005, Gheor92, Gheor2000} 
\begin{subequations}\label{quadyn39}
\begin{equation}\label{quadyn39a}
K_{0,1r} = \frac 1{2\hbar \omega_r}\left[\frac 1{2M} p_\rho^2 \pm \frac{M\omega_r^2}2 \rho^2 + \frac{\hbar^2}{2M}\left(l^2 - \frac 14\right) \frac 1{\rho^2}\right] \ ,
\end{equation}
\begin{equation}\label{quadyn39b}
K_{2r} = \frac i4\left[2\rho \frac{\partial}{\partial \rho} + 1\right] \ ,
\end{equation}
\end{subequations}
where $\omega_r/2 \pi$ is the trap radial frequency and $p_\rho = -i\hbar \frac \partial {\partial \rho}$. Because the angular momentum operator commutes both with the generators of the symplectic groups and the quantum Hamilton function $H$, we can restrict the study of the quantum system to a subspace of the Hilbert space \cite {Busch2016} with axial angular momentum $L_z = \hbar l I$, where $l$ represents the orbital quantum number \cite{Mihalcea2011}. The commutation relations are 
\begin{multline}\label{quadyn44}
\left[K_{0r}, K_{1r}\right] = iK_{2r} \ ,\;\left[K_{0r}, K_{2r}\right] = -iK_{1r} \ , \;
\left[K_{1r}, K_{2r}\right] = iK_{0r} \ 
\end{multline}
Using eq. (\ref{quadyn39}) we infer
\begin{equation}\label{quadyn54}
K_{0r}^2 - K_{1r}^2 = \frac 14\left(-\rho^2\frac{\partial^2}{\partial \rho^2} + l^2 - \frac 14\right)  + iK_{2r} . 
\end{equation}

The Casimir operator for the radial symplectic group states is \cite{Major2005, Perel86, Gheor92, Mihalcea2011}
\begin{equation}\label{quadyn55}
C_{2r} = K_{0r}^2 - K_{1r}^2 - K_{2r}^2 = \frac{l^2-1}4 = k\left(k-1\right) \ ,
\end{equation}
therefore 
\begin{equation}\label{quadyn56}
k_{r12} = \frac{1\pm l}2\Rightarrow k_r = \frac{l+1}2 ,\; k_r > 0 \; \mbox{and} \; l \geq 0 \ .
\end{equation}

We choose a potential with axial symmetry \cite{Major2005, Mihalcea2011}
\begin{equation}\label{quadynEP}
 \Phi\left(\vec r, t\right) = A(t)g\left(\rho, z\right) \ ,
 \end{equation}
where $A\left(t\right)$ is a time periodic function, of period $T = 2\pi / \Omega$, where $\Omega$ is the frequency of the RF trapping voltage, and $g$ is a function of $\rho^2$ and $z^2$
\begin{equation}\label{pot}
 g\left(\rho, z\right) = \sum_{k \geq 1} c_k H_{2k}\left(\rho, z\right) \ .
\end{equation}
The $H_{2k}$ polynomials are harmonic, of  degree $k$ in $\rho^2$ and $z^2$. The electric potential described by eq. (\ref{quadynEP}) is characteristic to a trap that exhibits both axial and radial (with respect to the $xOy$ plan) symmetry. In case of a Penning trap $A$ is constant in time, but for most RF (Paul) traps $A\left(t\right) = U_0 + V_0\cos{\Omega t}$, where $U_0$ is the d.c. voltage and $V_0$ is the RF trapping voltage. In case of harmonic potentials $c_k = 0 \ , \mbox{for}\; k>1$
\begin{equation}
 g\left(\rho, z\right) = \frac 1{r_0^2 + 2 z_0^2} \left(\rho^2 - 2z^2\right) \ , \; c_2 = - \frac 1{r_0^2 + 2 z_o^2} \ ,
\end{equation}
where $r_0$ and $z_0$ represent the semiaxes of the combined quadrupole trap and $\rho^2 = x^2 + y^2$. The particular case of an ideal Paul trap is obtained for zero magnetic field. In case of a RF ion trap with octupole anharmonicity, the electric potential given by eq. (\ref{pot}) becomes 
\begin{multline}
 g\left(\rho, z\right) = c_1 H_2\left(\rho, z\right) + c_2 H_4 \left(\rho, z\right) + \ldots \\
 H_2\left(\rho, z\right) = \frac 12\left(2 z^2 - \rho^2\right) \ , \; H_4\left(\rho, z\right) = \frac 18\left(8 z^4 - 24 z^2 \rho^2 + 3 \rho^4\right) \ ,
\end{multline}
where $H_2k$ represent the harmonic polynomials \cite{Major2005}. We denote the elastic constants as 
\begin{equation}
 K_r = \frac{M\omega_c^2}4 - 2 Qc_2 A(t) \ , \; K_a = 4Qc_2 A(t) \ .
\end{equation}

The quantum Hamilton function that describes an ion (particle) of electric charge $Q$, mass $M$, and orbital angular momentum $\hbar l$, confined within an octupole ion trap characterized by a RF anharmonic electric potential, can be expressed as \cite{Mihalcea2011}
\begin{multline}\label{quadyn57}
H_l = -\frac{\hbar^2}{2M}\left(\frac{\partial^2}{\partial \rho^2} - \frac{l^2}{\rho^2} + \frac 1\rho \frac \partial {\partial \rho}\right) -  \frac{\hbar^2}{2M}\frac{\partial^2}{\partial z^2}+\frac{K_r}2\rho^2 + \frac{K_a}2z^2 - \\ \frac{\omega_c}2\hbar l + QA\left(t\right) P\left( \rho^2, z^2\right) \ ,
\end{multline}
where $\omega_c = QB_0/M$ stands for the cyclotronic frequency, and $B_0$ is the axial magnetic field. The anharmonic part can be expressed as  
\begin{equation}
P\left(\rho^2, z^2\right) =\sum_{k\geq 2}c_kH_{2k}\left(\rho, z\right) \;,
\end{equation}
where $c_k$ are constants \cite{Mihalcea2011}, and $H_{2k}$ are harmonic polynomials (spherical harmonics functions) of $k$ degree in $\rho^2$ and $z^2$, defined as
\begin{equation}
 H_{2k}\left(\rho, z\right) = \sum_{j = 0}^k \frac{\left(2k\right)!\rho^{2j}z^{2(k-j)}}{4^j \left(2k - 2j\right)!\left(j!\right)^2} \ .
\end{equation}

In order to characterize the quantum system, we must build the Hamilton function. We investigate the Schr\"odinger equation
\begin{equation}
 i\hbar \frac{\partial \chi}{\partial t} = H_l \chi \ , 
\end{equation}
where the quantum Hamilton function $H_l$ results from eq. (\ref{quadyn57}), particularized for combined quadrupole or octupole traps. Thus, the quantum Hamilton function describes an algebraic model when $P\left(\rho^2, z^2\right)$ is a polynomial function. Such model is linear for quadrupole traps ($P$ is a linear combination of $\rho^2$ and $z^2$). We choose $\mu = -l^2$ and we reintroduce the Lie algebra generators for the radial symplectic group 
\begin{subequations}\label{quadyn58}
\begin{equation}\label{quadyn58a}
K_{0,1r}^{\prime} = \frac 1{2\hbar \omega_r}\left[\pm \frac{\hbar^2}{2M}\left( \frac{\partial^2}{\partial \rho^2} + \frac{\mu}{\rho^2} + \frac 1\rho \frac \partial {\partial \rho}\right) + \frac{M\omega_r^2}2\rho^2\right] \ ,
\end{equation}
\begin{equation}\label{quadyn58b}
K_{2r}^{\prime} = \frac i2\left( \rho \frac \partial {\partial \rho} + 1\right) \ ,
\end{equation}
\end{subequations}
and 
\begin{equation}\label{quadyn59}
\left[K_{0r}^{\prime}, K_{1r}^{\prime}\right] = iK_{2r}^{\prime} \ ,\;\left[K_{2r}^{\prime}, K_{1r}^{\prime}\right] = iK_{0r}^{\prime}\ ,\\
\left[ K_{2r}^{\prime}, K_{0r}^{\prime}\right] = iK_{1r}^{\prime} \ .
\end{equation} 

Using eqs. (\ref{quadyn58}) we infer
\begin{subequations}\label{quadyn65}
\begin{equation}\label{quadyn65a}
\rho^2 = \frac{2\hbar}{M\omega_r}\left(K_{0r}^{\prime} + K_{1r}^{\prime}\right) \, 
\end{equation}
\begin{equation}\label{quadyn65b}
- \frac{\hbar^2}{2M}\left(\frac{\partial^2}{\partial \rho^2} + \frac{\mu}{\rho^2} + \frac 1\rho\frac \partial {\partial \rho}\right) = \hbar \omega_r\left(K_{0r}^{\prime} - K_{1r}^{\prime}\right) \ .
\end{equation}
\end{subequations}
Using eqs. (\ref{quadyn37}), the quantum Hamilton function described by eq. (\ref{quadyn57}) changes accordingly
\begin{multline}\label{quadyn67}
H_l = \hbar \omega_r\left( K_{0r}^{\prime} - K_{1r}^{\prime}\right) + 2\hbar \omega_a\left(K_{0a} - K_{1a}\right) + \frac{K_r}2\frac{2\hbar }{M\omega_r}\left(K_{0r}^{\prime } + K_{1r}^{\prime}\right) + \\ 
\frac{K_a}2\frac{2\hbar}{M\omega_a}\left(K_{0a} + K_{1a}\right) - \frac{\omega_c}2\hbar l + QA\left(t\right) P\left(\rho^2, z^2\right) \ .
\end{multline}
We consider our trap to exhibit axial symmetry (in the coordinates $\rho$ and $z$) and rotation symmetry with respect to the $xOy$ plan. From eqs. (\ref{quadyn59}) and (\ref{quadyn65}) we infer
\begin{equation}\label{quadyn68}
\left(K_{0r}^{\prime} + K_{1r}^{\prime}\right)\left(K_{0r}^{\prime} - K_{1r}^{\prime}\right) = C_2 + K_{2r}^{\prime 2} - iK_{2r}^{\prime} \ ,
\end{equation}
and the Casimir operator is 
\begin{equation}
\label{quadyn66}C_{2r} = -\frac 14\left(\mu + 1\right) \Rightarrow k_r = \frac{l + 1}2 \ ,\; \mbox{with} \; l \geq 0 \ .
\end{equation}
Hence, the study of the Hamilton function for an ion trapped within the combined (or Paul) trap we have considered is reduced to the study of a linear Hamilton system for the real symplectic group $Sp\left( 2,{\mathbb R}\right)$. Then
\begin{equation}
P\left(\rho^2, z^2\right) = DH_4\left(\rho, z\right) = D\left(8z^4 - 24z^2\rho^2 + 3\rho^4\right) 
\end{equation}
where $D$ stands for a coefficient that depends on the trap geometry. We perform an average on the generators of the symplectic group, and on $\rho^4, \rho^2z^2$ and $z^4$. The anharmonic term is 
\begin{equation}
\Phi_{anh} = A\left(t\right) DH_4\left(\rho, z\right) \ ,
\end{equation}
while in the quasipotential approximation case (ideal Paul trap), the anharmonic term can be expressed as
\begin{multline}
\Phi_{anheff} = C_4H_4\left(\rho, z\right) + C_6H_6\left(\rho, z\right) \ , \\
 H_6\left(\rho, z\right) = \frac 1{16} \left(-5 \rho^6 + 90 \rho^4 z^2 - 120 \rho^2 z^4 + 16 z^6\right) 
\end{multline}
where the $C_4$ and $C_6$ coefficients depend on the specific trap geometry. Hence, the quantum Hamilton function can be expressed as 
\begin{equation}
\label{quadyn70}H_l = H_{anh} + \left\{
\begin{array}{ll} Q\Phi_{anh} & \mbox{Paul and Penning trap,}\\ 
Q\Phi _{anheff} &  \, \mbox{ideal Paul trap}  \ .
\end{array} 
\right\}
\end{equation}
Eq. (\ref{quadyn70}) gives the quantum Hamilton function for the particular traps we have chosen, including the pseudopotential approximation case that characterizes an ideal Paul trap.   

\paragraph{Schr\"odinger equation solutions and energy spectrum}\label{energy}
The solutions of the Schr\"odinger equation for the Hamilton function described by eq. (\ref{quadyn67}) are \cite{Major2005, Gheor2000}
\begin{equation}
 \Psi_{k_am_ak_rm_rl} = \frac 1{\sqrt \rho} \exp \left[il\left(\theta + \frac{\omega_c}2 t\right) - i\varphi \right]\psi_{k_am_a}\left(z_a\right)\psi_{k_rm_r}\left(z_r\right) \;,
\end{equation}
where $m_a, m_r \in {\mathbb N}$ are natural numbers, $z_a, z_r \in {\mathbb Z}$ are complex coordinates within the unit disk $\left(|z_a| < 1, \;|z_r| < 1\right)$, and $\varphi \in {\mathbb R}$ represents a real phase such as
\begin{equation}
 \varphi = \left(k_a + m_a\right)\varphi_a + \left(k_r, m_r\right)\varphi_r \;.
\end{equation}
The $z_a, z_r, \varphi_a$ and $\varphi_r$ variables are solutions of the differential equations
\begin{equation}
 i\frac{dz_j}{dt} = \alpha_j + \frac{\beta_j}2\left(z_j^2 + 1\right) \;,\; \frac{d\varphi_j}{dt} = \alpha_j + \frac{\beta_j}2\left(z_j + z_j^*\right)\;,\; j = a, r.
\end{equation}
The functions $\psi_{k_am_a}\left(z_a\right)$ and $\psi_{k_rm_r}\left(z_r\right)$ represent the coherent symplectic vectors, for symplectic axial and radial dynamic groups. In order to explicitly define these vectors, we recall the orthonormal system of coherent states associated to the group SU(1,1) \cite{Nemoto2000}. We use the raising and lowering operators $K_{\pm j}$ given by eq. (\ref{quadyn71}), and the canonical base \cite{Perel86, Gheor2000}  
\begin{equation}
 \phi_{k_jm_j} = \left[\frac{\Gamma\left(2k_j\right)}{m_j! \Gamma\left(2 k_j + m_j\right)}\right]^{1/2} \left(K_{+j}\right)^{m_j} \phi_{k_j 0} \;,
\end{equation}
where the normalized vector satisfies $K_{+j}\phi_{k_j 0} = k_j\phi_{k_c 0}$ and $K_{-j} \phi_{k_j 0} = 0$. We  introduce the unitary evolution operator defined as  
\begin{equation}
 U\left(z_j\right) = \exp\left(z_jK_{+j}\right)\exp\left(\lambda_jK_{0j}\right)\exp\left(-z_j^*K_{-j}\right) \;,
\end{equation}
where $\lambda_j = \ln (1- z_jz_j^*)$. The coherent symplectic vectors are defined as
\begin{equation}
 \psi_{k_jm_j}\left(z_j\right) = U\left(z_j\right) \phi_{k_jm_j} \;.
\end{equation}
The quasienergy spectrum is thus determined by the $\Psi_{k_am_ak_rm_rl}$ wavefunctions and the quasienergies \cite{Gheor2000}
\begin{equation}
 E_{k_am_ak_rm_rl} = 2\hbar \left[\mu_a\left(k_a + m_a\right) + \mu_r\left(k_r + m_r\right) - \frac{\omega_cl}4\right] \;,
\end{equation}
where $\mu_a$ and $\mu_r$ represent the Floquet exponents for the stable solutions of the Hill equations. 

The quantum Hamilton function for a system of $N$ ions confined within a cylindrically symmetric RF quadrupole ion trap is found in \cite{Gheor2000}. The center of mass (CM) Hamilton function is similar to the single particle Hamiltonian given in \cite{Mihalcea2011}, which means that the quasienergy and coherent states are almost the same (the single ion mass is substituted with the product of masses in case of the multi-ion system). When the interaction potential is invariant to translations and homogeneous of rank $-2$, soluble models result (e.g., for Calogero type potentials). Such models enable obtaining explicit bases and systems of coherent states in order to investigate trapped ion systems \cite{Combe2012}. 

\section{Quasiclassical dynamics in nonlinear combined and Paul traps}\label{quasi}
\paragraph{Classical Hamilton function for an anharmonic octupole and quadrupole trap. Equations of motion}

We recall some definitions with respect to coherent states built over weight states of discrete positive series of the symplectic group $Sp\left(2, {\mathbb R}\right)$ \cite{Dragt2005}. The Lie algebra associated to this group is spanned by  three generators, namely $K_0, K_1$, and $K_2$, where the raising and lowering operators $K_{\pm} = K_1\pm iK_2$, obey the following commutation relations \cite{Major2005, Combe92, Gheor92}
\begin{equation}
\label{quadyn71}\left[K_0, K_{\pm}\right] = \pm K_{\pm}\ ,\;\ \left[K_{-}, K_{+}\right] = 2K_0 \ , 
\end{equation}
The eigenvalues of the Casimir operator are denoted by $k\left(k-1\right) $, where $k$ stands for the Bargmann index for unitary irreducible representations (UIR) \cite{Dragt2005}. In such case we denote the basis vector as $\left| m,k\right\rangle ; m = 0, 1,\ldots $, where $K_0$ is diagonal. The action of the operators $K_+$ and $K_-$ is defined in \cite{Gheor92, Dragt2005}. We introduce the following squeezed coherent states for the group $Sp\left(2, {\mathbb R}\right)$:
\begin{equation}
\label{quadyn74}\left|z, m, k\right\rangle = U\left(z\right) \left|m, k\right\rangle \ ,
\end{equation}
where the unitary evolution operator for the quantum Hamiltonian (which satisfies the Schr\"odinger equation) is \cite{Combe92}
\begin{equation}
\label{quadyn75}U\left(z\right) = \exp\left(zK_+\right)\exp\left(\lambda K_0\right) \exp\left(-z^* K_-\right) \ ,
\end{equation}
are the operators corresponding to the group representation with $\left|z\right| < 1$ and $\lambda = \ln\left(1 - z z^*\right)$. For $m = 0$, the geometric construction of Perelomov results \cite{Perel86, Gheor92}. The matrix elements for the generators $K_+, K_-$, and $K_0$ in the squeezed coherent states $\left| z,k,m\right\rangle$ can be expressed as \cite{Gheor92, Mihalcea2011}  
\begin{equation}
\label{quadyn78}
\begin{array}{lll}
\left\langle z, k, m\right|K_+\left|z, k, m\right\rangle = 2\left(k + m\right)z^*\left(1 - zz^*\right)^{-1} &  &  \\ 
\left\langle z, k,m \right|K_-\left|z, k, m\right\rangle = 2\left(k + m\right)z\left(1 - zz^*\right)^{-1} &  &  \\ 
\left\langle z, k, m\right|K_0\left|z, k, m\right\rangle = \left(k + m\right)\left(1 + zz^*\right) \left(1 - z z^*\right)^{-1} &  &  \ .
\end{array}
\end{equation}

We denote $\Omega = K_0 + K_1 = K_0 + \frac 12\left(K_{+} + K_{-}\right)$. Then
\begin{equation}
\label{quadyn80} U^\dagger \left(z\right) \Omega U\left(z\right) = \frac{\left(1 + z\right)\left(1 + z^*\right)}{\left(1 - zz^*\right)}\left(2K_0 + e^{-i\varphi}K_{-} + e^{i\varphi}K_{+}\right) \ ,
\end{equation}
where $U^\dagger$ represents the Hermitian adjoint of the unitary operator $U\left( z\right)$, and  
\begin{equation}
\label{quadyn81}e^{i\varphi} = \frac{1 + z}{1 + z^*}\ ,\; e^{-i\varphi} = \frac{1 + z^*}{1 + z}\ ,\; \varphi \in {\mathbb R} \ .
\end{equation}
Moreover
\begin{multline}\label{quadyn82}
  \left[U^\dagger \left(z\right)\Omega U\left(z\right) \right]^n = U^\dagger \left(z\right) \Omega^nU\left( z\right)  = \frac{\left(1 + z\right)^n\left(1 + z^*\right)^n}{\left(1 - zz^*\right)^n}E^n \ , \\
  E = 2K_0 + e^{-i\varphi}K_{-} + e^{i\varphi}K_{+} \ .
\end{multline}
From eqs. (\ref{quadyn78} - \ref{quadyn82}) we infer
\begin{equation}\label{quadyn821}
\left\langle z, k, m\right|\Omega^n\left|z, k, m\right\rangle = \frac{\left(1 + z\right)^n\left(1 + z^*\right)^n}{\left(1 - zz^*\right)^n} \left\langle k, m\right|E^n\left|k, m\right\rangle \ .
\end{equation}
Eq. \ref{quadyn821} allows the explicit calculus of the energy function for any dynamical group. By calculus we infer
\begin{multline}\label{quadyn83}
E^2 = 4K_0^2 + e^{-2i\varphi}K_-^2 + e^{2i\varphi}K_+^2 + K_- K_+ + K_+K_- + \\ 2e^{-i\varphi}\left(K_0K_{+} + K_{-}K_0\right) 
+ 2e^{i\varphi}\left(K_0K_{+} + K_{+}K_0\right) \ , 
\end{multline}
\begin{multline}\label{quadyn84}
 E^3 = 8K_0^3 + e^{-3i\varphi}K_{-}^3 + e^{3i\varphi}K_{+}^3 + 2e^{-2i\varphi}\left(K_{-}^2K_0 + K_0K_{-}^2 + K_{-}K_0K_{-}\right) \\ + 2e^{2i\varphi}\left(K_{+}^2K_0 + K_0K_{+}^2 + K_{+}K_0K_{+}\right) +  
4e^{-i\varphi}\left(K_0K_{-}K_0 + K_{-}K_0^2 + K_0^2K_{-}\right) + \\ 4e^{i\varphi}\left(K_0K_{+}K_0 + K_{+}K_0^2 + K_0^2K_{+}\right) + 
e^{i\varphi}\left(K_+^2K_- + K_-K_+^2 + K_+K_-K_+\right) + \\
e^{-i\varphi}\left(K_{-}K_{+}K_{-} + K_{+}K_{-}^2 + K_{-}^2K_{+}\right) + \\
 2\left(K_0K_{+}K_{-} + K_{+}K_0K_{-} + K_{-}K_{+}K_0 + K_{+}K_{-}K_0 + K_0K_{-}K_{+}+ K_{-}K_0K_{+}\right) \ . 
\end{multline}

We weight on the vacuum state vector $\left| 0\right\rangle $ and obtain
\begin{equation}\label{quadyn85}
K_{-}\left|0\right\rangle = 0\ , \left\langle 0|0\right\rangle = 0 \ , K_0\left|0\right\rangle = k\left| 0\right\rangle \ , \left\langle 0\right| K_{+} = 0 \ . 
\end{equation}
Eq. (\ref{quadyn71}) leads to 
\begin{multline}\label{quadyn87}
  K_{-}K_{+} = K_{+}K_{-} + 2K_0 \ , K_{-}K_0 = K_0K_{-} + K_{-} \ , \\ 
  K_0K_{+} = K_{+}K_0 + K_{+} \ .
\end{multline}

Then, the Casimir operator can be expressed as
$$
C_2 = K_0^2 - \frac 14\left(K_{+} + K_{-}\right)^2 - \frac 1{4i^2}\left(K_{+} - K_{-}\right)^2 = k\left(k - 1\right) \ , 
$$
and
\begin{equation}
\label{quadyn88}K_{+}K_{-} + K_{-}K_{+} = 2K_0^2 - 2k\left(k - 1\right) \ .
\end{equation}
Using eqs. (\ref{quadyn85}) we infer
\begin{subequations}\label{quadyn89}
\begin{equation}\label{quadyn89a}
\left\langle E^2\right\rangle = \left\langle 0|E^2|0\right\rangle = 4k^2 + 2k \ ,
\end{equation}
\begin{equation}\label{quadyn89b}
\left\langle E^3\right\rangle =\left\langle 0|E^3|0\right\rangle = 8k^3 + 12k^2 + 4k \ .
\end{equation}
\end{subequations}
If instead of the vacuum state vector $\left| 0\right\rangle $ we weight on a vector of an orthonormal system of state vectors $\left| k,m\right\rangle$, namely $\left\langle k, m|k, m^{\prime}\right\rangle = \delta_{mm^{\prime}}$, we obtain 
\begin{subequations}\label{quadyn90}
\begin{equation}\label{quadyn90a}
 \left\langle k, m|E^2|k, m\right\rangle = 2k\left(2k + 1\right) + 12km + 6m^2  \ ,
\end{equation}
\begin{equation}\label{quadyn90b}
\left\langle k, m|E^3|k, m\right\rangle = 4k\left(k + 1\right)\left(2k + 1\right) + 4mk\left(5 + 12k\right)  + 4m^2\left(15k + 1\right) + 20m^3 \ ,
\end{equation}
\end{subequations}
where $K_0\left|k, m\right\rangle = \left(k + m\right)\left|k, m\right\rangle$ \cite{Gheor92} and we make use of eqs. (\ref{quadyn85}) and (\ref{quadyn89}). Using eq. (\ref{quadyn87}), we are able to evaluate the final term from eq. (\ref{quadyn84}) as
\begin{multline}\label{quadyn010}
F = 2K_0\left(K_{+}K_{-} + K_{-}K_{+}\right) + 2\left(K_{+}K_{-} + K_{-}K_{+}\right)K_0 + \\
2K_0\left(K_{+}K_{-} + K_{-}K_{+}\right) + 2\left(K_{+}K_{-} + K_{-}K_{+}\right) \ .
\end{multline}
We denote 
\begin{equation}
\label{quadyn91} \xi = \frac{\left(1 + z\right)\left(1 + z^*\right)}{1 - zz^*} = S_n \ ,\;n=1,2,\ldots \ ,
\end{equation}
where
\begin{subequations}\label{quadyn92}
\begin{equation}\label{quadyn92a}
S_1 = 2\xi \left(k + m\right) = S_1\left(z, k, m\right)
\end{equation}
\begin{equation}\label{quadyn92b}
S_2 = \xi^2\left[2k\left(2k + 1\right) + 12km + 6m^2\right]
\end{equation}
\begin{equation}\label{quadyn92c}
S_3 = \xi^3\left[4k\left(k + 1\right)\left(2k + 1\right) + 4mk\left(5 + 12k\right) + 4m^2\left(15k + 1\right) + 20m^3\right] \ .
\end{equation}
\end{subequations}
By taking into account eq. (\ref{quadyn78}) \cite{Major2005, Gheor92} and 
\begin{equation}
\label{quadyn93}\left\langle z, k, m|A|z, k, m\right\rangle = \left\langle k, m|U^{-1}\left(z\right) AU\left(z\right) |k, m\right\rangle \ ,
\end{equation}
we infer 
\begin{equation}
\label{quadyn94}\left\langle z, k, m|K_0 - K_1|z, k, m\right\rangle = \left(k + m\right)\frac{\left(1 - z\right) \left(1 - z^*\right) }{1 - zz^*} \ ,
\end{equation}
where $z$ stands for the squeezed state parametre, $k$ and $m$ represent pure harmonic oscillator states, while $K_0 - K_1$ is the kinetic energy. The classical energy (Hamilton) function for an anharmonic octupole ion trap described by eq. (\ref{quadyn67}), can be expressed as 
\begin{multline}\label{quadyn95}
 {\mathcal H}_{cl} = \hbar\omega_r\left(k_r + m_r\right)\frac{\left(1 - z_r\right)\left(1 - z^*_r\right)}{1 - z_rz^*_r} + 2\hbar \omega_a\left(k_a + m_a\right)\frac{\left(1 - z_a\right)\left(1 - z^*_a\right)}{1 - z_az^*_a} + \\
 \frac{2\hbar K_r}{M\omega_r} \left(k_r + m_r\right)\frac{\left(1 - z_r\right)\left(1 - z^*_r\right)}{1 - z_r z^*_r} + \frac{2\hbar K_a}{M\omega_a} \left(k_a + m_a\right)\frac{\left(1 - z_a\right)\left(1 - z^*_a\right)}{1 - z_a z^*_a} + H_{anh} - \frac{\omega_c}2\hbar l \ ,
\end{multline}
where 
\begin{equation}
 H_{anh} = Q A\left(t\right)\langle P\left(\rho^2, z^2\right) \rangle = QA\left(t\right)\sum_{k\geq 1}c_k\langle H_{2k}\rangle .
\end{equation}
We have denoted by $\langle X\rangle$ the expectation value of the $X$ operator in the coherent state $\Phi_{m_am_r}\left(z_a, z_r\right)$. In particular we obtain
\begin{equation}\label{quadyn96}
H_{anh} = \left\{ 
\begin{array}{ll}
QA\left(t\right)D\left\langle H_4\right\rangle \ , \mbox{combined (Paul and Penning) trap}\ , &  \\ 
QC_4\left\langle H_4\right\rangle + QC_6\left\langle H_6\right\rangle , \mbox{ideal Paul trap} & \ .
\end{array}
\right. 
\end{equation}
The second term in eq. (\ref{quadyn96}) characterizes the pseudopotential aproximation case (the ideal radiofrequency or Paul trap) \cite{Major2005}. Then 
\begin{equation}
\label{quadyn97}\left\langle H_4\right\rangle = 8S_{2a} - 24S_{1r}S_{1a} + 3S_{2r} \ ,
\end{equation}
\begin{equation}
\label{quadyn98}\left\langle H_6\right\rangle = 16S_{3a} - 120S_{2a}S_{1r} + 90S_{1a}S_{2r} - 5S_{3r} \ ,
\end{equation}
\begin{equation}
\label{quadyn99)}H_6 = 16 z^6 - 120 z^4\rho^2 + 90 z^2\rho^4 - 5\rho ^6 \ .
\end{equation}
We denote
$$
S_{jr} = S_j\left(z_r, k_r, m_r\right) \ \ ,\;\;S_{ja} = S_j\left(z_a, k_a, m_a\right) \ \ ,\;\;1\leq j\leq 3 \ ,
$$
and then introduce the Husimi $Q$ representation
\begin{equation}\label{quadyn100}
S_j\left(z, k, m\right) = \xi^jQ_j\left(k, m\right) \ .
\end{equation}
Moreover
\begin{subequations}\label{quadyn101}
\begin{equation}\label{quadyn101a}
Q_1\left(k, m\right) = 2\left(k + m\right)
\end{equation}
\begin{equation}\label{quadyn101b}
Q_2\left(k, m\right) = 2k\left(2k + 1\right) + 12km + 6m^2
\end{equation}
\begin{equation}\label{quadyn101c}
Q_3\left(k, m\right) = 4k\left(k + 1\right)\left(2k + 1\right) + 4mk\left(5 + 12k\right) + 4m^2\left(15k + 1\right) + 20m^3 \ .
\end{equation}
\end{subequations}
The parametre values are known
$$
k_a=\frac 14, \frac 34\ ;\;\;m_a, m_r = 0, 1,\ldots \;;\;\;k_r = \frac{l + 1}2 \ ,
$$
where $l$ stands for the orbital quantum number for the trapped particle (ion), and the orbital angular momentum is $L^2 = l\left(l + 1\right)I$. By minimizing ${\mathcal H}$ after $z$ we infer the approximations for the quantum energies. Using cylindrical coordinates $\left(z = \rho e^{i\theta }\ ,\; z^* = \rho e^{-i\theta }\right)$, we introduce the complex variables 
\begin{equation}\label{quadyn102}
\xi_n = \frac{\left(1 + z\right)\left(1 + z^*\right) }{1 - zz^*} \ ,\; \eta_n  = \frac{\left(1 - z\right)\left(1 - z^*\right)}{1 - zz^*}\ ,\; n = a, r \ .
\end{equation}

The classical Hamilton function for an anharmonic octupole trap can be expressed as 
\begin{multline}\label{quadyn104}
{\mathcal H}_{cl} = \hbar \omega_r\left(k_r + m_r\right)\eta_r + 2\hbar \omega_a\left(k_a + m_a\right)\eta_a + \frac{2\hbar K_r}{M\omega_r}\left(k_r + m_r\right) \xi_r + \\ \frac{2\hbar K_a}{M\omega_a}\left(k_a + m_a\right)\xi_a - \frac{\omega_c}2\hbar l 
\\
 +\left\{ 
\begin{array}{ll}
QA\left(t\right)D\left[8S_{2a} - 24S_{1r}S_{1a} + 3S_{2r}\right] \longmapsto \mbox{combined trap,}&  \\ 
QC_4\left(8S_{2a} - 24S_{1r}S_{1a} + 3S_{2r}\right) + \\ QC_6\left(16S_{3a} - 120S_{2a}S_{1r} + 90S_{1a}S_{2r} - 5S_{3r}\right) \longmapsto \mbox{pseudopot. approx.} \ ,
\end{array}
\right.
\end{multline}
where the first expression refers to the case of an octupole combined Paul and Penning trap, while the second case describes the pseudopotential approximation in case of an ideal Paul trap. The energy function associated to the quantum Hamilton function $H_l$ is a classical type Hamiltonian $H_{cl}$, whose values are exactly the expectation values of $H_l$ on the coherent states $\psi_{k_a 0}(z_a)$ and $\psi_{k_r 0}(z_r)$. Therefore
\begin{equation}
\label{quadyn105}\frac \partial {\partial z} = \frac{\partial \xi}{\partial z}\frac\partial{\partial \xi} + \frac{\partial \eta}{\partial z}\frac \partial {\partial \eta}\;\ ,\;\;\frac \partial {\partial z^*} = \frac{\partial \xi}{\partial z^*}\frac \partial {\partial \xi} + \frac{\partial \eta }{\partial z^*}\frac \partial {\partial \eta} \ .
\end{equation}
Quantum dynamics for ions confined in quadrupole and octupole combined and ideal Paul traps can be characterized by solutions of the Schr\"odinger equation, where the $H = H_r$ Hamilton function corresponds to radial motion, while $H = H_{a}$ describes the axial motion. We apply the TDVP and obtain the classical Hamilton function for the particular traps we considered   
$$
{\mathcal H_{cl}} = \left\langle z, k, m|H_l|z, k, m\right\rangle \ ,
$$
which determines an equation of motion in the classical Lobacevski phase space $\left| z\right|< 1$ :
\begin{equation}
\label{quadyn106}\dot z = \left\{z, {\mathcal H_{cl}}\right\} \ , 
\end{equation}
where $\left\{, \right\}$ stands for the generalized Poisson bracket. We finally obtain
\begin{equation}
\label{quadyn108}\left\{z, {\mathcal H_{cl}}\right\} = \frac{\left(1 - z z^*\right)^2}{2i\left(k + m\right)}\frac{\partial{\mathcal H_{cl}}}{\partial z^*} \ ,
\end{equation}
which represents the equation of motion for an ion confined within an octupole combined (or Paul) trap. Thus, we show that the coherent state approach leads to quantum solutions that are explicitly constructed as functions of the classical trajectories on the phase space.

\paragraph{Hamilton function for an anharmonic combined trap. Equations of motion}

According to eq. (\ref{quadyn104}) the classical Hamilton function can be expressed as
\begin{multline}\label{quadyn109}
{\mathcal H}_{cl\ anh} = \hbar\omega_r\left(k_r + m_r\right)\eta_r + 2\hbar\omega_a\left(k_a + m_a\right)\eta_a + \frac{2\hbar K_r}{M\omega_r}\left(k_r + m_r\right)\xi_r +  \\
\frac{2\hbar K_a}{M\omega_a}\left(k_a + m_a\right) \xi_a - \frac{\omega_c}2\hbar l + QA\left(t\right) D\left[8S_{2a} - 24S_{1r}S_{1a} + 3S_{2r}\right] \ .
\end{multline}

The expressions for $S_{1a}, S_{2a}, S_{1r}$ and $S_{2r}$ result from eqs. (\ref{quadyn92a}). We turn to the expression of the Hamilton function described by eq. (\ref{quadyn109}). The points of minimum of the Hamilton function are characterized by the following equations
\begin{subequations}\label{quadyn111}
\begin{multline}\label{quadyn111a}
  \frac{\partial {\mathcal H}_{lanh}}{\partial \xi_a} = 0\Rightarrow \frac{2\hbar K_a}{M\omega_a}\left(k_a + m_a\right) + \\ QA\left(t\right) D\left[16\xi _a\cdot Q_2\left(k_a, m_a\right) - 96\xi _r \left( k_a + m_a\right) \left(k_r + m_r\right)\right] = 0
\end{multline}
\begin{multline}\label{quadyn111b}
  \frac{\partial {\mathcal H}_{lanh}}{\partial \xi _r} = 0\Rightarrow \frac{2\hbar K_r}{M\omega_r}\left(k_r + m_r\right) + \\ QA\left(t\right) D\left[-96\xi_a\left(k_a + m_a\right)\left(k_r + m_r\right) + 6\xi_rQ_2\left(k_r, m_r\right) \right] = 0
\end{multline}
\begin{equation}\label{quadyn111c}
  \frac{\partial {\mathcal H}_{lanh}}{\partial\eta_a} = 0\Rightarrow 2\hbar\omega_a\left(k_a + m_a\right) = 0
\end{equation}
\begin{equation}\label{quadyn111d}
  \frac{\partial {\mathcal H}_{lanh}}{\partial \eta_r} = 0\Rightarrow 2\hbar\omega_r\left(k_r + m_r\right) = 0 \ .
\end{equation}
\end{subequations}
Using eq. (\ref{quadyn102}), we infer 
\begin{equation}\label{quadyn114}
1-zz^* = 2 \frac{1 + zz^*}{\xi + \eta} \ ,\; z + z^* = \frac 12\left(1 - zz^*\right)\left(\xi - \eta\right) \ . 
\end{equation}
According to eq. (\ref{quadyn105}) and eqs. (\ref{quadyn102}) 
\begin{equation}
\label{quadyn117}\frac{\partial {\mathcal H}}{\partial z^*} = \frac{\partial{\mathcal H}}{\partial \xi}
\left(\frac{1 + z}{1 - z z^*}\right)^2 - \frac{\partial{\mathcal H}}{\partial \eta}\left(\frac{z - 1}{1 - z z^*}\right)^2 
\end{equation}
From eqs. (\ref{quadyn102}) and (\ref{quadyn114}), after some calculus, we obtain
\begin{subequations}\label{quadyn119}
\begin{equation}\label{quadyn119a}
\dot\xi = \left(\frac{1 + z^*}{1 - zz^*}\right)^2\dot z + \left(\frac{1 + z}{1 - zz^*}\right)^2 z^* \ .
\end{equation}
\begin{equation}\label{quadyn119b}
\dot\eta = -\left(\frac{1 - z^*}{1 - zz^*}\right)^2\dot z - \left(\frac{1 - z}{1 - zz^*}\right)^2 z^* \ . 
\end{equation}
\end{subequations}

From eq. (\ref{quadyn106}), we can express $z^*$ as: 
\begin{equation}
\label{quadyn121} z^* = -\frac{\left(1 - zz^*\right)^2}{2i\left(k + m\right)}\frac{\partial {\mathcal H}}{\partial z} \ ,
\end{equation}
while eqs. (\ref{quadyn105}) yields
\begin{equation}
\label{quadyn122}\frac{\partial {\mathcal H}}{\partial z} = \frac{\partial {\mathcal H}}{\partial \xi}\left(\frac{1 + z^*}{1 - zz^*}\right)^2 - \frac{\partial {\mathcal H}}{\partial \eta}\left(\frac{z^* - 1}{1 - zz^*}\right)^2 \ .
\end{equation}
If we replace eqs. (\ref{quadyn106}) and (\ref{quadyn121}) in eq. (\ref{quadyn119}) and do the math, we infer 
\begin{equation}\label{quadyn124}
2i\left(k + m\right) \dot \xi = 4\frac{z - z^*}{1 - zz^*}\frac{\partial {\mathcal H}}{\partial \eta} \ . 
\end{equation}
If the same mathematical operations are applied to eq. (\ref{quadyn119b}), the result is
\begin{equation}
\label{quadyn126}2i\left(k + m\right)\dot\eta = - 4\frac{z - z^*}{1 - zz^*}\frac{\partial{\mathcal H}}{\partial \xi} \ . 
\end{equation}
We return to eq. (\ref{quadyn102}) and then write eq. (\ref{quadyn114}) as
\begin{equation}
\label{quadyn127}\frac 2{1 - zz^*} = \frac{\xi + \eta }2 + 1 \ .
\end{equation}
Then 
\begin{equation}
\label{quadyn128}z + z^* = 2\frac{\xi - \eta}{\xi + \eta + 2} \ ,\; zz^* =  \frac{\xi + \eta - 2}{\xi + \eta + 2} \ . 
\end{equation}
Using eqs. (\ref{quadyn128}) we infer
\begin{equation}
\label{quadyn130)}\left(z - z^*\right)^2 = 16 \frac{1- \xi \eta}{\left(\xi + \eta + 2\right)^2} \ .
\end{equation} 
Hence
\begin{equation}
\label{quadyn131}\frac{z - z^*}{1 - zz^*} = i\varepsilon \sqrt{\left|1 - \xi\eta \right|}\;\;,\;\varepsilon =\pm 1 \ .
\end{equation}
Considering eq. (\ref{quadyn131}), the equations of motion (\ref{quadyn124}) and (\ref{quadyn126}) can be expressed as 
\begin{subequations}
\begin{equation}\label{quadyn132a}
2\left(k + m\right) \dot \eta = -4\varepsilon \sqrt{\left|1 - \xi \eta \right| }\frac{\partial {\mathcal H}}{\partial \xi} 
\end{equation}
\begin{equation}\label{quadyn132b}
2\left( k + m\right) \dot \xi = 4\varepsilon \sqrt{\left| 1 - \xi \eta \right| }\frac{\partial {\mathcal H}}{\partial \eta } \ .
\end{equation}
\end{subequations}
Further on, we denote
\begin{equation}
i\sigma =\frac{z - z^*}{1 - zz^*}\Rightarrow -\sigma^2 = 1 - \xi\eta \;, 
\end{equation}
while the time derivative is
\begin{equation}
\label{quadyn134}2 \sigma \dot\sigma = \dot\xi \eta + \xi \dot\eta \ .
\end{equation}
Then eqs. (\ref{quadyn132a}) and (\ref{quadyn132b}) can be expressed as 
\begin{equation}\label{quadyn135}
\left(k + m\right) \dot\eta = -2\sigma\frac{\partial {\mathcal H}}{\partial \xi} \ , \; \left(k + m\right) \dot\xi = 2\sigma \frac{\partial {\mathcal H}}{\partial \eta}\;.
\end{equation}
By comparing eqs. (\ref{quadyn135}) and (\ref{quadyn134}), we infer the following equations of motion
\begin{equation}
\label{quadyn136}\left(k + m\right) \dot\sigma = \eta\frac{\partial{\mathcal H}}{\partial\eta } - \xi \frac{\partial {\mathcal H}}{\partial \xi} \ .
\end{equation}

\paragraph{Quasipotential approximation case - Ideal RF (Paul) trap case}

The expression of the classical Hamilton function for an octupole trap, in case of the pseudopotential approximation is (see eq. (\ref{quadyn104})) 
\begin{multline}
{\mathcal H}_{cl\ anheff} = \hbar\omega_r\left(k_r + m_r\right)\eta_r + 2\hbar\omega_a\left(k_a + m_a\right)\eta_a + \frac{2\hbar K_r}{M\omega_r}\left(k_r + m_r\right)\xi_r  + \\
 \frac{2\hbar K_a}{M\omega_a}\left(k_a + m_a\right) \xi_a - \frac{\omega_c}2 \hbar l + QC_4<H_4> + QC_6<H_6>
\end{multline}
where $H_4$ and $H_6$ are characterized by eqs. (\ref{quadyn97} - \ref{quadyn98}). Hence
\begin{equation}
 S_{3a} = 4k_a\left(k_a + 1\right)\left(2k_a + 1\right) + 4k_a m_a\left(5 + 12 k_a\right) + 4m_a^2 \left(15 k_a + 1\right) + 20 m_a^3 \ ,
\end{equation}
\begin{equation}
 S_{3r} = 4k_r\left(k_r + 1\right)\left(2k_r + 1\right) + 4k_r m_r\left(5 + 12 k_r\right) + 4m_r^2 \left(15 k_a + 1\right) + 20 m_r^3 \ .
\end{equation}
The points of minimum that characterize the equilibrium configurations for trapped ions are given by the equations
\begin{multline}\label{quadyn140}
  \frac{2\hbar K_a}{M\omega_a}\left(k_a + m_a\right) + QC_4\left[16\xi_a Q_2\left(k_a, m_a\right) - 96\xi_r \left(k_a + m_a\right) \left(k_r + m_r\right)\right] + \\
  QC_6 [48\xi_a^2 Q_3\left(k_a, m_a\right) - 240\xi_a \xi_r Q_2\left(k_a, m_a\right) Q_1\left(k_r, m_r\right) +  
  90\xi_r Q_1\left(k_a, m_a\right) Q_2\left(k_r, m_r\right)] = 0
\end{multline}
\begin{multline}\label{quadyn141}
  \frac{2\hbar K_r}{M\omega_r}\left(k_r + m_r\right) + QC_4 \left[-96\xi_a\left(k_a + m_a\right)\left(k_r + m_r\right) + 6\xi_r Q_2\left(k_r, m_r\right) \right] + \\
  QC_6[-120\xi_a^2 Q_2\left(k_a, m_a\right) Q_1\left(k_r, m_r\right) + 180\, \xi_a \xi_r Q_1\left(k_a, m_a\right) Q_2\left(k_r, m_r\right) - 15\xi_r^2 Q_3\left(k_r, m_r\right)] = 0 \ .
\end{multline}

In the pseudopotential approximation case, the points of minimum of the dequantified Hamilton function define the equilibrium configurations for trapped ions (Coulomb crystals), of interest for implementing and scaling quantum logic (see \cite{Marci2010}). The equilibrium configurations are found by performing numerical simulations, which is verified for example in case of a 2D octupole Paul trap, where  axially symmetric structures of Coulomb crystals and phase transitions have been reported \cite{Okada2009, Marci2012}. The advantage of a linear RF octupole trap lies in a large region within the trap centre where the electric potential is flat, which makes it very suited for cold molecule spectroscopy or frequency metrology \cite{Champ2010}. Moreover, experiments performed with higher order multipole 2D RF ion traps show the occurrence of multiple regions of dynamical stability, some of them located near the trap electrodes where the electrical potential is steep. Numerical simulations performed validate experimental data and observations \cite{Mihalcea2016a}. Recent investigations are focused on producing planar Coulomb crystals using specially designed linear multipole traps \cite{Wine2011, Okada2009, Champ2013}. Calculations performed for time-dependent dynamics in RF ion traps were also generalized to multispecies ion crystals in multipole traps \cite{Landa2012a}. 
       
The possibility to perform precision numerical simulations of the associated (semi)classical or quantum dynamics for any kind of engineered quantum system is an issue of large importance, and the methods presented in \cite{Singer2010} can be directly adapted to any given Hamiltonian, including the model we propose. Numerical simulations can be used to validate the method suggested in the paper and find the equilibrium configurations for trapped ions or the eigenvalues of the normal modes of oscillation \cite{Champ2010}. The results in the paper are also valid for different quantum simulator architectures \cite{Yoshi2015}. 

\section{Conclusions}\label{Concl}

The paper characterizes the evolution of squeezed coherent states for quadrupole and octupole RF and combined traps, an issue not studied in the literature. We propose a dequantization algorithm by which the energy function associated to the trapped ion system is a classical Hamilton function, whose values represent the expectation values of the quantum Hamiltonian on squeezed coherent states. Such formalism can be applied to Hamiltonians that are nonlinear in the infinitesimal generators of a dynamical symmetry group, such as the case of quadrupole and octupole ion traps with a RF anharmonic electric potential, which we demonstrate in the paper. An algorithm is proposed, that extends the results to any electrical multipole. We build the quantum Hamilton function associated to an ion confined within a nonlinear combined (or RF) ion trap. Moreover, we also build the classical Hamilton functions particularized for such traps. The classical (Hamilton) equations of motion in complex coordinates are obtained for an ion confined in an octupole combined trap. The results obtained are also valid in case of the center of mass (CM) motion, for a system made of identical ions.

To summarize, the coherent state formalism has been developed, particularized and applied in case of quadrupole and octupole RF and combined traps. We show that the dynamic group associated to an ion confined within such particular traps is the product of the symplectic axial group, a symplectic radial group and the group of rotations around the symmetry axis. The study of the Hamilton function for an ion trapped in an anharmonic combined (or RF) trap we have considered is reduced to the study of the linear Hamilton system for the real symplectic group $Sp\left( 2,{\mathbb R}\right)$. In agreement with the TDVP applied on coherent states, if a harmonic analytical electric potential is added, the classical equations of motion turn into coupled nonlinear Hill equations. We also show that the coherent state approach leads to quantum solutions that are explicitly constructed as functions of the classical trajectories on the Lobacevski phase space. 

Dequantization by means of coherent states enables achieving exact connections between quantum and classical dynamical systems, that exhibit (axial) symmetry. For an ion confined within a quadrupole ion trap (QIT), the requantization operation corresponds to a geometrical quanzation. The method(s) and formalism suggested are developed and implemented for the particular traps we considered. The points of minimum of the dequantified Hamilton function can be identified through standard numerical programming. The algorithm we suggest, combined with the latest experimental techniques, can be applied to investigate entanglement, quantum logic, and quantum simulations in a wide range of physical systems.

\section{acknowledgements}
The author would like to acknowledge support provided by the Ministery of Education, Research and Inovation from Romania (ANCS-National Agency for Scientific Research), contracts PN 16470103, UEFISCDI 90/06.10.2011, and ROSA Contract Nr. 53 / 20.11.2013 - LEOPARD.  

The author (B. M.) is grateful to Prof. Marian Apostol (IFIN-HH) and Prof. G\"{u}nther Werth for very useful comments and suggestions on the paper. 

\section*{References}

\bibliography{QuaDyn2}

\end{document}